\documentclass[notitlepage,aps,prd,amsmath,floats,floatfix,twocolumn,superscriptaddress,nofootinbib]{revtex4}

\usepackage{amsfonts,amsmath,wasysym,epsfig,graphicx,verbatim,color,subfigure,graphicx}
\usepackage[inkscapelatex=false]{svg}
\DeclareUnicodeCharacter{2009}{\,}

\usepackage[range-phrase=--,range-units=single,retain-unity-mantissa=false]{siunitx}
\usepackage{stmaryrd}
\usepackage{amsmath}
\usepackage{amssymb}
\usepackage{amsfonts}
\usepackage{bm}
\usepackage{color}
\usepackage[normalem]{ulem}
\usepackage{xcolor}
\usepackage{natbib}
\usepackage{color,soul}
\usepackage{hyperref}
\usepackage{marginnote}
\usepackage{braket}
\usepackage[none]{hyphenat}
\usepackage{bm}
\usepackage{comment}
\usepackage{multirow}
\usepackage{xspace}
\usepackage[toc,page]{appendix}
\usepackage{float}
\usepackage[acronym]{glossaries}
\usepackage{blkarray}

\newcommand*{\COMMENTS}{} 

\ifdefined\COMMENTS

\else

\fi

\newcommand{\SU}{\affiliation{Department of Physics, Syracuse University, Syracuse, New York 13244, USA}}

\usepackage[utf8]{inputenc}
\usepackage[T1]{fontenc}

\newacronym{PDH}{PDH}{Pound-Drever-Hall}
\newacronym{HG}{HG}{Hermite-Gaussian}
\newacronym{LG}{LG}{Laguerre-Gaussian}
\newacronym{NBI}{NBI}{Neutral Beam Injection}
\newacronym{QPD}{QPD}{Quadrant Photo-Detector}
\newacronym{BPD}{BPD}{Bullseye Photo-Detector}

\date{\today}

\hypersetup{
    colorlinks=true,
    linkcolor=blue,
    filecolor=magenta,      
    urlcolor=cyan,
    pdftitle={Overleaf Example},
    pdfpagemode=FullScreen,
    }

\begin{document}

\title{Combined Cavity Alignment and Mode-Mismatch Sensing using RF-QPD Sensors}

\author{Mitchell Schiworski}\email{mitchell.schiworski@ligo.org}\SU
\author{Stefan Ballmer}\SU

\begin{abstract}
In this article we describe a scheme for sensing both the mode mismatch and alignment of a cavity which does not require custom sensors and uses only \glspl{QPD}.
The technique is simple relying upon cylindrical lenses and thoughtful Gouy phase telescope design, and is trivial to implement alongside a \gls{PDH} locking scheme.
The Gouy phase telescope design utilizes all three of the \glspl{QPD} to sense each degree of freedom which also increases sensitivity and provides redundancy.
The technique we show also works without RF modulation, which is useful for providing other general diagnostic information in complex optical experiments.
An analytical overview of the technique as well as generic design equations for the Gouy phase telescope is provided.
Finally, a detailed optical simulation is presented to verify the technique.
\end{abstract}

\maketitle
\section{Introduction}\label{sec:introduction}
The most basic requirement for achieving resonance of an optical cavity is matching the optical round-trip length and the wavelength of the resonating light.
Typically this requires an active control scheme which actuates accordingly on the cavity length or laser frequency, the most ubiquitous of which being the \gls{PDH}~\cite{drever_laser_1983,black_introduction_2001} method.
Achieving an optimal resonance of the cavity, where the input light maximally resonates in the desired cavity mode however, requires consideration to the precise alignment of the input beam/cavity mirrors and to the size/shape of the input beam wavefront.
These three conditions we refer to as the length, alignment and mode-matching of the cavity.
In the best case scenario, misalignment and mode-mismatch decrease the coupling of the incident light into the cavity reducing the resonating power.
In the worst case, it may significantly seed higher order cavity modes which co-resonate with the fundamental mode.
This can degrade cavity length sensing in \gls{PDH} schemes preventing locking altogether or altering the feedback loop gain and introducing offsets that reduce the lock stability and performance.

For the typical tabletop experiment, active control of the alignment and mode-matching of the cavity is not required.
Gravitational wave detectors however, use suspended cavity optics necessitating active alignment controls~\cite{morrison_automatic_1994}.
Likewise, the high cavity powers result in significant heating of the cavity mirrors via optical absorption.
This creates effective thermal lenses in mirror substrates and thermal expansion alters the curvature of the cavity mirrors~\cite{hello_analytical_1990,hello_analytical_1990-1,vinet_special_2009}, both of which cause mode mismatch which needs to be controlled and compensated for~\cite{lawrence_active_2003,rocchi_thermal_2012,brooks_overview_2016}.
Even before such thermal effects arise, static mode mismatch from errors in the placement of optics, tolerances in curvatures/focal lengths and non-normal incidence on focusing optics all degrade detector performance.
Mode mismatch remains a pressing issue for gravitational wave interferometers and currently limits both the sensitivity and duty cycle~\cite{mcculler_ligos_2021,goodwin-jones_transverse_2024,kuns_squeezed_2026}.
Many other extreme optical power experiments are susceptible to thermally induced mismatch limiting the power which can be achieved.
This includes: photoneutralisation cavities for generating neutral ion beams used for \gls{NBI} in fusion reactors~\cite{simonin_negative_2016}, high harmonic generation of extreme-ultraviolet sources~\cite{pupeza_compact_2013}, generation of x-ray sources via Compton scattering~\cite{jacquet_first_2024} and the generation of gamma ray sources at CERN~\cite{martens_design_2022}.
Any experiment that requires active mode-matching generally also requires a complementing alignment scheme.
Mode-matching actuators, such as lenses on translation stages or mirrors/lenses with adjustable focal lengths, will deflect the beam as well as actuate on the wavefront unless one is aligned directly through the center of curvature.
It is practically impossible to completely eradicate this effect on the cavity alignment.

The article is structured as follows: next, in Section~\ref{sec:background} we give a background on cavity alignment \& modematching, followed by an overview of wavefront sensing schemes used to sense cavity misalignment and modematching.
In Section~\ref{sec:trident} we present a scheme for combined alignment and modematching sensing based on these principles requiring only three \glspl{QPD} and no custom sensors.
Finally, in Section~\ref{sec:simulation}, we show results from a detailed optical simulation comparing the error signals from this technique against are more conventional \gls{QPD} and annular segmented sensing scheme.
\section{Background}\label{sec:background}
\subsection{Cavity alignment and mode-matching}
\begin{figure}
    \centering
    \includegraphics[]{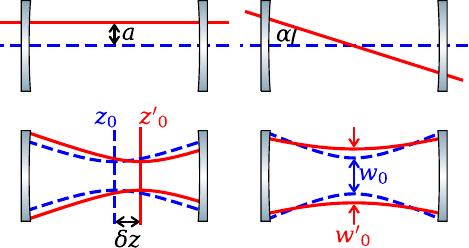}
    \caption{(Top) Cavity misalignment degrees of freedom of translation $a$ and tilt $\alpha$ of the input optical axis relative to the cavity axis. (Bottom) Cavity mismatch degrees of freedom of waist position $\delta z$ and waist size $\delta w_0$ errors.}
    \label{fig:cav_misalignment_mismatch}
\end{figure}

Unlike the length condition, alignment and mismatch each have two orthogonal degrees of freedom~\cite{anderson_alignment_1984}.
Following the treatment in Ref.~\cite{anderson_alignment_1984}, for cases of misalignment it can be projected in terms of a \emph{translation} or \emph{tilt} of the input axis relative to the cavity axis for each transverse dimension.
For mode-mismatch it can be projected in terms of an error in the input beam \emph{waist size} and \emph{waist position} relative to the cavity fundamental mode.
These cases are both summarized in Fig.~\ref{fig:cav_misalignment_mismatch}.
Misalignments of the cavity mirrors (rather than the input beam) can be represented as a combination of translation and tilt errors via geometric relations~\cite{anderson_alignment_1984}.
Likewise, substrate thermal lensing on the input mirror and thermal expansion induced radius of curvature changes in the mirrors can be represented as a combination of waist size and position errors via ABCD calculations.

A misaligned or mismatched incident beam can be represented as some perfectly matched/aligned beam containing extra higher order \gls{HG} modes.
Some incident Gaussian beam of frequency $\omega$ described by some mismatched basis $\tilde{q}'$ traveling along some misaligned axis $\hat{r}'$, then can be represented as:
\begin{equation}\label{eq:e_o_expansion}
    HG_{00}(\tilde{q}',\hat{r}')\sqrt{P}e^{i\omega t} = \sqrt{P}\sum_{n,m=0}^{\infty}\beta_{nm}HG_{nm}(\tilde{q_c},\hat{r})e^{i\omega t}
\end{equation}
where $\omega$ is the frequency, $\tilde{q}_c$ is the cavity basis and $\hat{r}$ is aligned to the cavity axis.
For small misalignment and mismatch defined according to Fig.~\ref{fig:cav_misalignment_mismatch}, the mode coefficients $\beta_{nm}$ for $n+m\leq2$ are approximately~\cite{anderson_alignment_1984}:
\begin{equation}\label{eq:alignment_coupling}
    \beta_{10} = a_x/w_0 + i\alpha_x\frac{\pi w_0}{\lambda}, \,\, \beta_{01} = a_y/w_0 + i\alpha_y\frac{\pi w_0}{\lambda}
\end{equation}
\begin{equation}\label{eq:mismatch_coupling}
    \beta_{20}=\beta_{02} = 1-w'_{0}/w_0 + i\frac{\delta z}{2\pi w_0^2}
\end{equation}
\begin{equation}
    |\beta_{00}|^2 = 1-\sum_{nm}|\beta_{nm}|^2
\end{equation}
In other words, misalignment of the input beam appears as first order \gls{HG} modes and mismatch appears as second order \gls{HG} modes.
When the mismatch is not astigmatic as shown in Eq.~\ref{eq:mismatch_coupling} even amounts of $HG_{20}$ and $HG_{02}$ appear which is equivalent to a first order \gls{LG} mode.
The full expressions of these expansions are derived in detail in Ref.~\cite{bayer-helms_coupling_1984}.

\subsection{Differential wavefront sensing}
The \gls{PDH} technique can also be expanded for sensing cavity misalignment and mode-mismatch.
This general approach is referred to as differential wavefront sensing, since one measures the difference between the wavefront of the incident beam and the resonating cavity field.
This difference is sensed and nullified to maximize the coupling into the cavity.
This type of alignment sensing is well refined and used throughout gravitational wave interferometers~\cite{morrison_automatic_1994}.
The extension to mismatch sensing has been demonstrated with a variety of sensing methods~\cite{mueller_determination_2000,magana-sandoval_sensing_2019,brown_differential_2021}.
Other schemes which rely on similar principles, but don't explicitly measure the differential wavefront have also been explored~\cite{ciobanu_mode_2020,goodwin-jones_single_2023}.
However, none of these have yet been implemented within the interferometers.
A clear method for mismatch sensing which is both simple and robust has not yet emerged.
In this section we introduce a framework for understanding each of these techniques together, rather than separately as they are typically represented in the literature.

\begin{figure}
    \centering
    \includegraphics[]{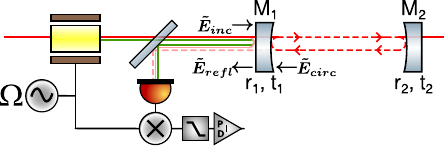}
    \caption{Generic differential wavefront sensing scheme.}
    \label{fig:dws_schematic}
\end{figure}

Fig.~\ref{fig:dws_schematic} shows a simplified schematic of a differential wavefront sensing scheme.
The cavity here is already assumed to be partially on resonance.
The phase modulation frequency $\Omega$ is chosen such that the sidebands are well outside the linewidth of the cavity and so are assumed anti-resonant.
A detector placed on reflection of the cavity is demodulated to isolate the component of the reflected field which oscillates at frequency $\Omega$.
The derivation in Appendix~\ref{sec:appendix_dws} shows that this signal component $S$ is proportional to:
\begin{equation}\label{eq:dws_signal}
\begin{split}
    S \propto\mathrm{Im} \bigg{[} \beta^*_{00}\kappa^*(\omega) HG^*_{00} (\tilde{q_c})
     \sum_{n,m=0}^\infty \beta_{nm} HG_{nm}(\tilde{q_c}) \bigg{]}
\end{split}
\end{equation}
where the $\beta_{nm}$ describe input field as a summation of \gls{HG} modes in the $\tilde{q_c}$ basis defined by the cavity and $\kappa(\omega)$ is a complex scalar which discriminates the laser wavelength vs the cavity length.
The terms $\beta_{nm}$ for $n,m>0$ describe the wavefront errors in the incident beam.

Different detector architectures are used to separately measure error signals proportional to $\kappa(\omega)$ and $\beta_{nm}$.
Let $M_i(x,y)$ be a mask function which describes the detector architecture, we can denote some error signal that detector would measure $E_i$ as:
\begin{equation}\label{eq:error_signal_form}
    E_i = \int_{-x_0/2}^{+x_0/2}\int_{-y_0/2}^{+y_0/2} S(x,y)M_i(x,y) \,dx\,dy
\end{equation}
where $x_0, y_0$ is the physical size of the detector in each dimension.
The mask $M_i(x,y)$ ideally has overlap with only the relevant $HG_{nm}$ mode.

\begin{figure}
    \centering
    \includegraphics[]{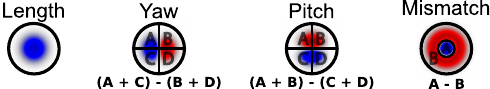}
    \caption{Different degrees of freedom for a cavity, the corresponding modes generated in the reflected signal and detector combinations which isolate these modes.}
    \label{fig:detector_masks}
\end{figure}

Fig. \ref{fig:detector_masks} summarizes the typical detector architectures to measure alignment and mode mismatch.
A single element photo-detector is used to measure the length degree of freedom.
When the detector is large relative to the beam then we have $M_{length}(x,y) = 1 \,\forall \,x,y$ giving:
\begin{equation}
    E_{length} = \int\int_{-\infty}^{+\infty} S(x,y) \,dx\,dy = |\beta_{00}|^2\mathrm{Im}[\kappa^*(\omega)]
\end{equation}
which in Appendix \ref{sec:appendix_dws} we show is equivalent to the solution derived in Ref.~\cite{black_introduction_2001}.
For alignment sensing, channels of a \gls{QPD} can be combined to extract the relevant coefficients $\beta_{10}$ and $\beta_{01}$.
For mismatch sensing, so-called \glspl{BPD} with annular segments have been created to extract the relevant mode mismatch error signals~\cite{mueller_determination_2000}.
Phase cameras~\cite{cao_optical_2019,agatsuma_high-performance_2019,muniz_high_2021} have also been demonstrated for alignment and mismatch sensing by processing the images with different digital summation masks~\cite{brown_differential_2021}.
The major drawback of these techniques is the custom sensors required which are less refined, more expensive and less sensitive compared to \glspl{QPD}.
\glspl{BPD} have different sized segments which make it difficult to precisely calibrate the gain of individual channels, which then introduces offsets in error signals.
Offsets also arise when the beam size on the detector is not optimally matched to the annular segments.
All phase camera designs have comparatively much lower bandwidth than \glspl{QPD}, and would require a more complicated form of digital processing to create error signals from the measured images.
\subsection{Accumulated Gouy phase shift}
Consideration of the accumulated Gouy phase~\cite{erden_accumulated_1997} of the beam reflected from the cavity to the detector(s) is not necessary for length sensing, however it is crucial for alignment and mismatch sensing.
In Eq.~\ref{eq:dws_signal} one can see that only the imaginary component of the expression appears in the detected signal $S$.
Physically this is because this quadrature results in an intensity modulation, while the other real quadrature is purely a phase modulation not measurable by the detection scheme.
At the same time, from Eqs.~\ref{eq:alignment_coupling} and ~\ref{eq:mismatch_coupling} we see the two degrees of freedom in misalignment/mismatch sensing are distinguished by the quadrature in which the modes appear.
Seemingly it is not possible to measure both degrees of freedom.
To get around this limitation one needs to use at least two detectors which each measure the reflected beam at different accumulated Gouy phases.

The accumulated Gouy phase shift $\phi$ of a $HG_{00}(\tilde{q})$ mode through an $ABCD$ optical system is given by~\cite{bond_interferometer_2017}:
\begin{equation}
    \exp(i\phi) = \left(\frac{A+B/\tilde{q}^*}{|A+B/\tilde{q}^*|} \right)
\end{equation}
and for a $HG_{nm}(\tilde{q})$ mode, the accumulated Gouy phase is $\exp(i\phi(1+n+m))$.
If Eq.~\ref{eq:dws_signal} describes the signal measured for a detector placed right at the input plane of the cavity, the signal measured by a detector at another axial location after some accumulated Gouy phase shift $\phi$ is:
\begin{equation}
\begin{split}
    S(\phi) \propto\mathrm{Im}& \bigg{[} \beta^*_{00}\kappa^*(\omega) HG^*_{00} (\tilde{q}')\\
     &\sum_{n,m=0}^\infty \beta_{nm} HG_{nm}(\tilde{q}')e^{i(n+m)\phi} \bigg{]}
\end{split}
\end{equation}
where we note that the $\tilde{q_c}$ is also transformed via the $ABCD$ system to $\tilde{q}'$.

As an example we can consider the case for misalignment in the horizontal axis, where for simplicity we look only at the first order modes.
From Eq.~\ref{eq:alignment_coupling}, our error signal measured by the detector looks like:
\begin{equation}\label{eq:gouy_example}
    \mathrm{Im} \left[ \left(a_x/w_0 + i\alpha_x\frac{\pi w_0}{\lambda} \right)e^{i\phi} \right]
\end{equation}
If one detector is placed such that $\phi=0+k\pi$ where $k$ is some integer, then the signal it measures contains only the $\alpha_x$ term relating to the tilt degree of freedom.
Another detector placed such that $\phi=\pi/2+k\pi$ contains only the $a_x$ term relating to the translation degree of freedom.
In reality, one only needs to control the difference in accumulated Gouy phase to each detector such that $\Delta\phi= \pi/2 +k\pi$.
Afterwards a calibration procedure then can be used to work out the correct linear combination of the signals in each detector which represents the individual degrees of freedom.
If $\Delta\phi \neq \pi/2$ both degrees of freedom can be sensed but not with equal SNR.
For mismatch sensing where the second order modes are relevant, the Gouy phase accumulates twice as fast and so the ideal separation is $\Delta\phi=\pi/4$.
\section{Three QPD mismatch and alignment sensing scheme}\label{sec:trident}
\begin{figure}
    \centering
    \includegraphics[]{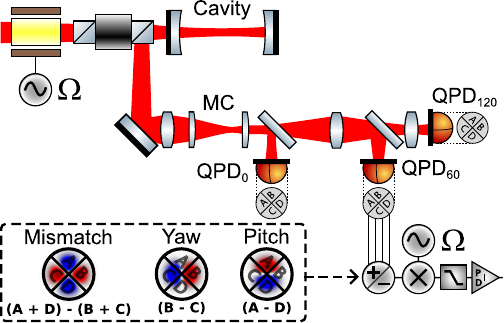}
    \caption{Schematic of the three \gls{QPD} mismatch and alignment scheme. The reflected cavity field is isolated and first sent through a mode converter (MC), then a Gouy phase telescope distributes the beam between the three \glspl{QPD} which have relative accumulated Gouy phases of $0^\circ,60^\circ\,$ and $120^\circ$. The output channels of the \glspl{QPD} are demodulated and combined as shown to create mismatch and alignment signals.}
    \label{fig:overview_schematic}
\end{figure}

Fig.~\ref{fig:overview_schematic} shows a generic overview of the proposed scheme.
In this section we describe individual elements of the technique in detail.
\subsection{Mode converter}
In order to circumvent the need for a custom detector design to resolve the 2nd order modes for mode-mismatch sensing, our proposed scheme implements a $\pi/2$ mode converter~\cite{oneil_mode_2000} in the path before the detectors.
These have previously been demonstrated for measuring mode-matching error signals with \glspl{QPD}~\cite{magana-sandoval_sensing_2019}.
After passing through the mode converter, the $LG_{10}(\tilde{q})$ mode is transformed into a $HG_{11}(\tilde{q})$ mode rotated by $45^\circ$, which is detectable using a \gls{QPD}.
Importantly, the $HG_{10}(\tilde{q})$ and $HG_{01}(\tilde{q})$ modes remain intact so alignment sensing is preserved.
\begin{figure}
    \centering
    \includegraphics[]{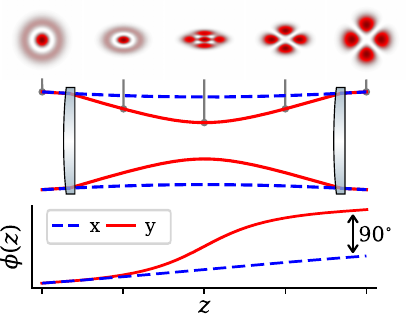}
    \caption{Intensity profile, beam spot size and accumulated Gouy phase of an $LG_{10}$ mode as it propagates through a mode converter.}
    \label{fig:mode_converter}
\end{figure}

The mode converter design consists of two cylindrical lenses as is shown in Fig.~\ref{fig:mode_converter}.
For one optical axis the lenses have no focusing power, whereas for the other optical axis the lenses focus the beam down through a waist and match the wavefront back to the non-focused axis.
After the mode converter, the beam is no longer astigmatic however the accumulated Gouy phase for the focused axis is now $\phi' = \phi+\pi/2$.
If we elect the y axis to be the focusing axis, then the accumulated Gouy phase after the mode converter then is given by:
\begin{equation}\label{eq:mode_converter_gouy}
    \exp \left[ i\left( 1+n+m\right)\phi \right]\exp \left[ i(1/2 + m)\pi/2 \right]
\end{equation}
where $\phi$ is the Gouy phase the non focusing axis accumulates through the mode converter. In appendix we show that for a $\phi'-\phi=\pi/2$ mode converter, we have the constraint $\phi=\pi/4$. We also show that  
Fig.~\ref{fig:mode_converter} lists the transformation of different modes through the mode converter.
To explain the transformation of the $LG_{10}(\tilde{q})$ mode, we first note the relation that:
\begin{equation}
    LG_{10}(\tilde{q}) = \frac{1}{\sqrt{2}}HG_{20}(\tilde{q}) + \frac{1}{\sqrt{2}}HG_{02}(\tilde{q})
\end{equation}
using Eq.~\ref{eq:mode_converter_gouy}, we can see that after the mode converter this becomes:
\begin{equation}
    \frac{1}{\sqrt{2}}HG_{20}(\tilde{q}) - \frac{1}{\sqrt{2}}HG_{02}(\tilde{q})
    = HG_{11}^{45^\circ}(\tilde{q})
\end{equation}
where have dropped the common phase factor of $e^{i\phi}$ and used the Euler identity $e^{i\pi}=-1$.
Fig.~\ref{fig:overview_schematic} shows the combinations of segments which create the alignment and mode-matching error signals for \gls{QPD} placed after the mode converter.
Note that it is necessary for either the \gls{QPD} to be rotated $45^\circ$, or equivalently the cylindrical lenses may be rotated instead.
From Fig.~\ref{fig:overview_schematic} it may appear that using only two of the segments of the \gls{QPD} to measure the alignment error signals results in reduced SNR, however this is not the case.
The situation is largely the same as the conventional sensing technique with a non-rotated \gls{QPD} with a rotated coordinate system.
The rotation however does slightly increase the susceptibility to cross talk between pitch \& yaw when the beam is not perfectly centered on the diode~\cite{kawabe_orientation_2006}.

\subsection{Gouy phase telescope design}
A key component of this sensing scheme is the use of only three detectors for detecting alignment and mode mismatch.
The three detectors are positioned within the same Gouy phase telescope such that they each contribute to mismatch and alignment sensing, increasing the shot noise sensitivity.

As stated in the previous section, the ideal Gouy phase separation of two detectors to measure alignment and mode mismatch is $\pi/2$ and $\pi/4$ respectively.
Here we derive the ideal Gouy phase separation of three detectors to measure both alignment and mode mismatch.
From Eq.~\ref{eq:gouy_example}, we can write that a detector at an accumulated Gouy phase $\phi$ from the cavity yields a signal which in general consists of a combination of both misalignment degrees of freedom:
\begin{equation}
    a_x/w_0\sin(\phi) + \alpha_x \frac{\pi w_0}{\lambda}\cos(\phi)
\end{equation}
In general it is not practical to have control over the actual accumulated Gouy phase of each detector, but rather the difference in their accumulated Gouy phase.
We then say that the three detectors are each placed at accumulated Gouy phases of $\phi$, $\phi+\Delta\phi_1$ and $\phi+\Delta\phi_2$ and treat $\phi$ as an uncontrolled variable.
The ideal placement of the detectors is one such that both the quadratures of translation and tilt are sensed with equal amplitudes.
Since the sign of the amplitude in which each quadrature is sensed is not relevant, we can write that the optimal $\Delta\phi_1, \Delta\phi_2$ fit the following constraint:
\begin{equation}\label{eq:alignment_gouy_phase_cond}
\begin{split}
    \sin^2(\phi)+\sin^2(\phi+\Delta\phi_1)+\sin^2(\phi+\Delta\phi_2) =\\
    \cos^2(\phi)+\cos^2(\phi+\Delta\phi_1)+\cos^2(\phi+\Delta\phi_2)
\end{split}
\end{equation}
which must hold true for any $\phi$.
This equation gives the solution that the ideal separation of the detectors is $\Delta\phi_1 = \pi/3 + k\pi$, $\Delta\phi_2 = 2\pi/3 + k\pi$.
Moreover we can generalize this proof that the ideal separation of $N$ detectors to measure alignment is:
\begin{equation}
    \Delta\phi_j = j\frac{\pi}{N}+k\pi, \,\,\, j=1\dots N{-}1
\end{equation}
We also require that the same is true for the two mode mismatch quadratures of waist size and waist position.
The equivalent solution for mismatch is:
\begin{equation}
    \Delta\phi_j = j\frac{\pi}{2N}+k\frac{\pi}{2}, \,\,\, j=1\dots N{-}1
\end{equation}
The solution we opt for that satisfies both conditions for three detectors is $\Delta\phi_1=\pi/3$, $\Delta\phi_2=2\pi/3$.

The Gouy phase telescope enforces the separation and importantly matches the beam width at each of the detectors.
This is a crucial part of any differential wavefront sensing scheme.
Appendix~\ref{sec:appendix_gouy_phase_telescope} presents the analytical solution for the telescope which can be easily adapted to an arbitrary cavity.
For this design the beam input to the telescope is assumed to be focused to a waist of size $w_{in}$.
One can adapt the solution for an arbitrary beam input by placing more focusing elements prior.
The other free parameter which must be specified is the desired beam spot size radius on the detectors, which we denote $w_{pd}$.
The values of $w_{in}$ and $w_{pd}$ are substituted into and Eq.~\ref{eq:gouy_telescope_1}-\ref{eq:gouy_telescope_3} to yield the parameters of the telescope shown in Fig.~\ref{fig:gouy_phase_telescope}.

\subsection{Sensing matrices and calibration}
Sensing matrices describe the combination of detector signals which give error signals for the two orthogonal degrees of freedom.
For two detectors optimally spaced in Gouy phase, the sensing matrices are simply rotation matrices.
In this scheme the mode converter and 3 detector Gouy phase telescope design mean the sensing matrices have different forms.
For alignment sensing in the non-focused axis of the mode converter we can write:
\begin{equation}
\begin{array}{c}
    M\\
    \begin{bmatrix}
    \sin(\phi) & \cos(\phi)\\
    \sin(\phi{+}\pi/3) & \cos(\phi{+}\pi/3) \\
    \sin(\phi{+}2\pi/3) & \cos(\phi{+}2\pi/3) \\
    \end{bmatrix}\\
\end{array}
\begin{array}{c}
    x\\
    \begin{bmatrix}
        a_x/w_0\\
        \alpha_x\frac{\pi w_0}{\lambda}\\
    \end{bmatrix}
\end{array}
=
\begin{array}{c}
    d\\
    \begin{bmatrix}
        \text{D}_{0}\\
        \text{D}_{60}\\
        \text{D}_{120}
    \end{bmatrix}
\end{array}
\end{equation}
where $D_i$ are the signals from each detector and $\phi$ is the accumulated Gouy phase from the cavity to the first detector.
The error signals $x$ are found by calculating the pseudo-inverse of the sensing matrix $M$:
\begin{equation}
    x=M^{-1}d
\end{equation}
For alignment sensing in the focused axis of the mode converter, the extra $\pi/2$ of accumulated Gouy phase gives:
\begin{equation}
M=
\begin{bmatrix}
    \cos(\phi) & {-}\sin(\phi)\\
    \cos(\phi{+}\pi/3) & {-}\sin(\phi{+}\pi/3) \\
    \cos(\phi{+}2\pi/3) & {-}\sin(\phi{+}2\pi/3)
\end{bmatrix}
\end{equation}
For mismatch sensing, after applying some trigonometric simplification we have:
\begin{equation}\label{eq:M_mismatch}
M=
\begin{bmatrix}
    \sin(2\phi) & \cos(2\phi)\\
    \sin(2\phi{+}2\pi/3) & \cos(2\phi{+}2\pi/3) \\
    -\sin(2\phi{+}\pi/3) & -\cos(2\phi{+}\pi/3)
\end{bmatrix}
\end{equation}
where $x=[w_0'/w_0-1,\, \delta z/z_R]^{\top}$.

One must perform some calibration procedure to measure the value of $\phi$.
In a control system one also needs to calibrate similar matrices that describe the effect of each actuator in terms of the two degrees of freedom.
Usually one skips the intermediate calibration and dithers the actuators and records how they appear in the corresponding detector channels.
However, in such cases as the LIGO interferometers where the mode-matching actuators operate on thermal timescales of multiple hours~\cite{brooks_overview_2016}, it would be cumbersome to do this calibration for the mode-mismatch sensing matrices.
A potential advantage of this sensing scheme then is that the alignment sensing matrix calibration can in theory be used to calibrate the mode-matching sensing matrix since the same detectors are used for both.
For a gravitational wave interferometer with coupled cavities, these relations are not straight forward to derive however and require further investigation.
\subsection{DC alignment and mode-matching sensing}
Another aspect of this sensing technique is worth highlighting: 
The DC signals from the \glspl{QPD} produce error signals  for alignment and mismatch. These error signals correspond to the alignment of the input beam relative to the telescope and the mode-matching of the beam into the mode converter.
The DC alignment signals are the same as any typical beam pointing setup using \glspl{QPD}.
Secondary DC alignment loops are a practical requirement for this kind of technique to ensure that the beam remains centered on the detectors.
Otherwise, spurious offsets and cross couplings are introduced as the detector segments work to cancel out unwanted modes.

With this technique one can also produce DC mode-matching error signals.
The mode converter defines a specific $\tilde{q}$ basis which an input Gaussian beam travels through and comes out non-astigmatic.
When the input beam is mismatched by this $\tilde{q}$, the output beam is astigmatic and produces signals for the mismatch channels of the detectors.
These correspond to waist size and waist position errors with Eq.~\ref{eq:M_mismatch} where $\phi$ is now the accumulated Gouy phase from the input of the mode converter to the first detector.

This DC version of the scheme may also be used to precisely characterize pointing noise and mode shape changes in lasers, or to infer material properties of optics by precisely measuring thermal lensing effects.
This differential readout is robust against intensity noise, and the use of \glspl{QPD} rather than cameras for measuring the mode shape lends towards a much higher sensitivity.
Even when employing the RF sensing scheme, these DC mismatch signals are useful for general diagnostics on the carrier beam in complicated experiments like gravitational wave interferometers.
In applications with extreme mismatch such that the beam size on the detector drastically changes or the accumulated Gouy phase of the detectors changes drastically, a separate mode-matching actuator may be required to combat clipping, spurious offsets and cross-couplings in the error signals.
The DC mismatch signals could be used to feed back to this actuator to ensure the $\tilde{q}$ of the carrier mode into the telescope remains constant.

\section{Optical simulation}\label{sec:simulation}
\begin{figure}
    \centering
    \includegraphics{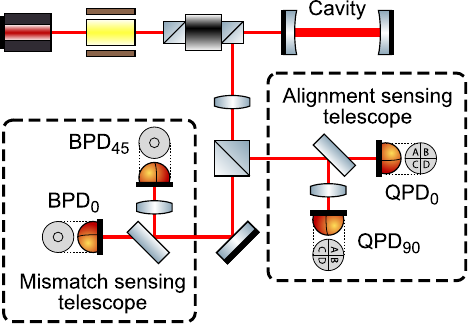}
    \caption{Schematic of the \textsc{Finesse} simulation used for comparison with the three \gls{QPD} technique. Using a more conventional method, the reflected beam is split into an alignment sensing scheme with two \glspl{QPD} with relative Gouy phases of $0^\circ$ and $90^\circ$, and a mismatch sensing scheme with two \glspl{BPD} with relative Gouy phases of $0^\circ$ and $45^\circ$.}
    \label{fig:simulation_schematics}
\end{figure}

To verify the technique we compare optical simulations of two techniques for sensing the alignment and mismatch of a cavity.
The cavity design mimics that of an aLIGO arm cavity, and the reflected beam is focused down to a waist of $w_{in}\approx0.8708$~mm input to the telescope and a spot size on the detectors of $w_{pd}=0.3$~mm.
The first simulation is as shown in Fig.~\ref{fig:overview_schematic}.
We compare this to the more conventional scheme in Fig.~\ref{fig:simulation_schematics} with two separate Gouy phase telescopes, one for alignment using \glspl{QPD} and one for mismatch using \glspl{BPD}.
The simulations were conducted using \textsc{Finesse}~\cite{brown_2025_12662017}, a frequency domain and HG modal based optical simulation package.
Modes of order up to $n+m\leq20$ were used.
The mode converter and Gouy phase telescope are implemented with basic lens components.
The detector signals are simulated by calculating the electric fields of the carrier and sidebands in cartesian co-ordinates and combining that into a differential wavefront error signal $S(x,y)$ according to Eq.~\ref{eq:beat_field}.
Corresponding detector masks $M(x,y)$ are generated to create error signals using Eq.~\ref{eq:error_signal_form}.
The radius of the inner segment used for the \gls{BPD} is 0.589 of the beam spot size on the detector when perfectly matched.
This value balances the total optical power on the two segments~\cite{bond_analytical_2016}.
This simulation method ensures that secondary effects such as higher order modes, changes to the beam spot size on the detectors, mismatch into the mode converter and changes to the detector accumulated Gouy phases are encapsulated.
A \gls{PDH} scheme is simulated to control the length of the cavity to maintain resonance and decouple the phase delay component of the induced misalignment and mismatch from the error signals.
\begin{figure}
    \centering
    \includegraphics[width=\linewidth]{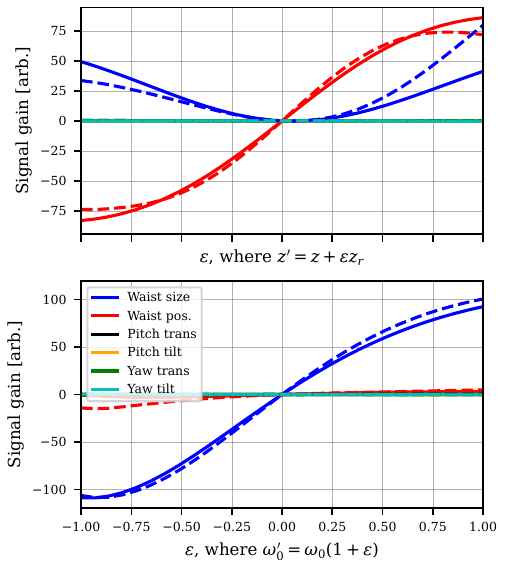}
    \caption{Error signals generated as the input beam is mismatched in the waist position degree of freedom (top) and the waist size degree of freedom (bottom). The solid lines correspond to the three detector mode converter technique in Fig.~\ref{fig:overview_schematic}, dashed lines correspond to the conventional technique shown in Fig.~\ref{fig:simulation_schematics}.}
    \label{fig:mismatch_results}
\end{figure}
\begin{figure}
    \centering
    \includegraphics[width=\linewidth]{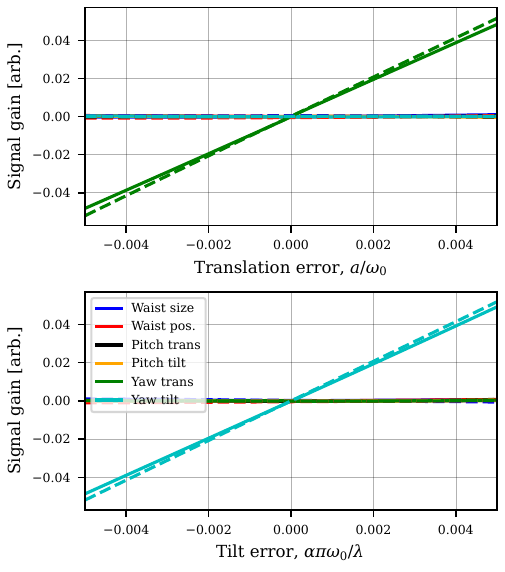}
    \caption{Error signals generated as the input beam is misaligned in the translation (top) and tilt (bottom) position degrees of freedom. The solid lines correspond to the three detector mode converter technique in Fig.~\ref{fig:overview_schematic}, dashed lines correspond to the conventional technique shown in Fig.~\ref{fig:simulation_schematics}.}
    \label{fig:alignment_results}
\end{figure}

The simulation results for mode mismatch are shown in Fig.~\ref{fig:mismatch_results}.
For small mismatches the results from the two techniques yield almost equivalent error signals.
For large mismatches a difference arises due to the different ways in which modes other than the $LG_{10}$ couple into the detectors.
The main culprit for these plots is the $LG_{20}$ terms which become significant for sufficiently large mismatch.
The bullseye detector architecture is comparatively more susceptible to the $LG_{20}$ mode which creates the discrepancy.

The simulation results for misalignment are shown in Fig.~\ref{fig:alignment_results}.
For these results only a smaller range of misalignments could be swept, this in part due to the cavity lock becoming unstable and to the way in which misalignments of beams are represented in the simulation.
In \textsc{Finesse}, the misalignment of a beam from the optical axis is represented by adding additional higher order modes.
Increasingly higher order modes are required in the simulation in order to remain physically accurate which exponentially increases the computational cost.
One can see that the error signals from the two results also agree other than a slight difference in the slope.
Unlike for mismatch sensing, the coupling of higher order modes through both techniques into the alignment channels is effectively the same as they both utilize \glspl{QPD}.

For the ease of comparison, the error signals in Fig.~\ref{fig:mismatch_results} and Fig.~\ref{fig:alignment_results} show the signal gain scaled to the light incident on a single detector in the scheme.
However, assuming there is adequate optical power to split between the detectors the overall sensitivity of the three \gls{QPD} scheme is higher by a factor of $\sqrt{3/2}$, owing to the fact that all three detectors contribute to each error signal.
In situations where the optical power is limited, the three detector scheme is also more efficient as each detector has more optical power available and all the optical power is used for sensing each degree of freedom.
\section{Conclusion}\label{sec:conclusion}
We have presented a technique for sensing mode mismatch and misalignment of a cavity which uses only \glspl{QPD} and does not require custom sensors.
By using 3 detectors for both mismatch and alignment sensing, the technique also offers higher sensitivity, redundancy if a detector becomes inoperable, and is more efficient in the use of optical power than a conventional technique using pairs of \glspl{QPD} and \glspl{BPD}.
We also have shown this technique to work simultaneously in the DC regime without the need for RF modulation, which is useful for general diagnostic purposes in complicated experiments.
The detailed optical simulation shows the error signals are equivalent to other methods.
\section{Acknowledgment}\label{sec:ack}
The work in this paper was supported by the National Science Foundation award 
PHY-2513058 and PHY-2309296.

\begin{appendices}
\section{Differential wavefront sensing error signal}\label{sec:appendix_dws}
The following derivation exists in more detail in Ref.~\cite{schiworski_development_2024}, we reproduce it here for consistency in notation with this article.
Referring to Fig.~\ref{fig:dws_schematic}, for small phase modulation index we can write the incident field $\tilde{E}_{inc}$ as~\cite{bond_interferometer_2017}:
\begin{equation}\label{eq:e_inc}
\tilde{E}_{inc} = E_0[ \sqrt{P_c}e^{i\omega t} - \sqrt{P_s}e^{i(\omega -\Omega)t} + \sqrt{P_s}e^{i(\omega +\Omega)t} ]
\end{equation}
where $E_0$ describes the non-temporal spatial component of the electric field, $P_c$ and $P_s$ are the powers in the carrier and each first order sidebands respectively, and $\omega$ is the carrier laser frequency.
We can represent $E_0$ as the summation of \gls{HG} modes in the basis defined by the cavity~\cite{siegman_lasers_1986}:
\begin{equation}\label{eq:e_o_expansion}
    E_0 = \sum_{n,m=0}^{\infty}\beta_{nm}HG_{nm}(\tilde{q_c})
\end{equation}
where $\beta_{nm}$ are complex mode coefficients normalized such that $\sum_{nm}|\beta_{nm}|^2=1$.
For most well behaved cavities, one can assume that only the fundamental mode of the carrier resonates in the cavity with significant power.
The portion of the resonating cavity mode which transmits through the input mirror and reaches the detector $\tilde{E}_{cav}$ is then ~\cite{siegman_lasers_1986}:
\begin{equation}\label{eq:e_circ}
    \tilde{E}_{cav} = \kappa(\omega)\sqrt{P_c}\beta_{00}HG_{00}(\tilde{q_c})
\end{equation}
the frequency dependent complex parameter $\kappa(\omega)$ is given by:
\begin{equation}
    \kappa(\omega) = \frac{-t_1^2 \tilde{g}_{rt}(\omega)}{r_1(1-\tilde{g}_{rt}(\omega))}
\end{equation}
with $\tilde{g}_{rt}(\omega)$ being the round trip gain for the $HG_{00}(\tilde{q_c})$ mode and $r_1,t_1$ denote the reflectivity transmissivity of the input mirror.
For modulation frequencies above the cavity pole, the sidebands are effectively anti-resonant and completely reflect off the cavity.
The reflected field also contains a component of the prompt reflection of the carrier field.
Combining this we get:
\begin{equation}\label{eq:e_refl}
\begin{split}
    \tilde{E}_{refl} =& \sqrt{P_c}\left[ r_1\sum_{n,m}\beta_{nm}HG_{nm}(\tilde{q_c}) \right. \\
    &+ \kappa(\omega) \beta_{00}HG_{00}(\tilde{q_c}) \bigg]e^{i\omega t} \\
    & + \sqrt{P_s}\sum_{n,m}\beta_{nm}HG_{nm}(\tilde{q_c})\left[ e^{i(\omega +\Omega)t} - e^{i(\omega -\Omega)t} \right]
\end{split}
\end{equation}

The detector measures a signal proportional to the intensity $|\tilde{E}_{refl}|^2$, which is then demodulated at the sideband frequency $\Omega$, isolating the components which oscillate at that frequency.
If we consider Eq.~\ref{eq:e_refl} in the form of:
\begin{equation}
    \tilde{E}_{refl} = Ae^{i\omega t} + B\left[  e^{i(\omega + \Omega)t} - e^{i(\omega - \Omega)t} \right]
\end{equation}
one can show that:
\begin{equation}
    |\tilde{E}_{refl}|^2 = |A|^2 + 2|B|^2 + 4\sin(\Omega t)\mathrm{Im}\left[ A^*B \right] + O(2\Omega)
\end{equation}
where $\mathrm{Im}$ denotes the imaginary operator.
The signal isolated by the demodulation circuit, which we denote the differential wavefront signal $S$ is then:
\begin{equation}\label{eq:beat_field}
    S \equiv \left. |\tilde{E}_{refl}|^2 \right|_{\Omega} = 4\mathrm{Im}[A^*B]
\end{equation}
substituting $A, B$ with the appropriate terms in Eq.~\ref{eq:e_refl}:
\begin{equation}
\begin{split}
    S =& 4\sqrt{P_cP_s}\mathrm{Im}\bigg[ \sum_{n,m}\beta_{nm}HG_{nm}(\tilde{q_c})\\
    & \bigg( r_1\sum_{n,m}\beta^*_{nm}HG^*_{nm}(\tilde{q_c}) + \kappa^*(\omega) \beta^*_{00}HG^*_{00}(\tilde{q_c}) \bigg)\bigg]
\end{split}
\end{equation}
The first product term can be discarded as it is the product of a complex expression with its conjugate and is hence purely real:
\begin{equation}
\begin{split}
    \mathrm{Im}\bigg[r_1\sum_{n,m}HG_{nm}(\tilde{q_c})\sum_{n,m}\beta^*_{nm}HG^*_{nm}(\tilde{q_c})\bigg] = 0
\end{split}
\end{equation}
which gives finally:
\begin{equation}
\begin{split}
    S =& \sqrt{P_cP_s}\mathrm{Im} \bigg{[} \beta^*_{00}\kappa^*(\omega) HG^*_{00} (\tilde{q_c})\\
     &\times\sum_{n,m} \beta_{nm} HG_{nm}(\tilde{q_c}) \bigg{]}
\end{split}
\end{equation}
For reassurance, we can take this equation and reproduce the \gls{PDH} error signals derived in Ref.~\cite{black_introduction_2001}.
A single element detector which is large relative to the beam diameter measures the quantity:
\begin{equation}
    \int\int_{-\infty}^{\infty} S\,dx\,dy = 4\sqrt{P_cP_s}|\beta_{00}|^2 \mathrm{Im}[\kappa(\omega)]
\end{equation}
where we have made use of the orthogonality property of the \gls{HG} modes:
\begin{equation}
    \int\int_{-\infty}^{\infty} HG^*_{nm}(\tilde{q})HG_{n'm'}(\tilde{q})\,dx\,dy = \delta_{nn'}\delta_{mm'}
\end{equation}
if we also assume no mode mismatch or misalignment then $|\beta_{00}|^2=1$.
We are then left with the equivalent \gls{PDH} error signal derived in Ref.~\cite{black_introduction_2001} for fast modulation frequencies since $\mathrm{Im}[\kappa(\omega)] = \mathrm{Im}[F(\omega)]$.

\section{Telescope design equations}\label{sec:appendix_gouy_phase_telescope}

\begin{figure}
    \centering
    \includegraphics[]{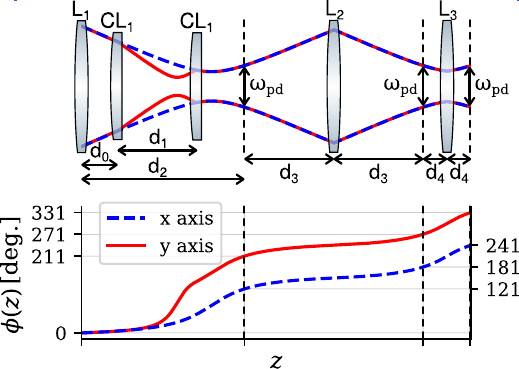}
    \caption{(Top) Beam size trace of the Gouy phase telescope design. (Bottom) Accumulated Gouy phase trace through the telescope. The vertical dashed lines indicate the positions of the detectors.}
    \label{fig:gouy_phase_telescope}
\end{figure}
\subsection{Mode converter}
\label{app:MC}
The design equations for a $\pi/2$ mode converter with two identical cylindrical lenses are given in Ref.~\cite{magana-sandoval_sensing_2019,beijersbergen_astigmatic_nodate}.
This symmetric design would not allow the mode converter to be folded within the Gouy phase telescope so we provide the design equations for a mode converter for some beam parameter $\tilde{q}_{mc}=z_{mc}+iz_{R_{mc}}$ at the input.

The distance between the two cylindrical lenses $d_{mc}$ must be the distance that the non-focused axis accumulates $\pi/4$ Gouy phase.
This arises from a fundamental relation one can derive that there are uniquely only two beams that can have the same width $\omega_a$ at one location $z_a$, and the same width $\omega_b$ at another location $z_b = z_a + d$.
It also follows that the accumulated Gouy phase $\phi$ for each of these beams over that distance $d$ are related by: 
\begin{equation}\label{eq:phi_sum}
    \phi+\phi'=\pi
\end{equation}
For the mode converter, we also have the requirement that the difference in the accumulated Gouy phase between the two axes is $\pi/2$ which gives:
\begin{equation}\label{eq:phi_diff}
    \phi' - \phi = \pi/2
\end{equation}
One can see that the only solution which satisfies Eq.~\ref{eq:phi_sum} and Eq.~\ref{eq:phi_diff} is $[\phi,\phi'] = [\pi/4,3\pi/4]$.
Therefore, $d_{mc}$ must be the distance that the non-focused axis accumulates $\pi/4$ Gouy phase which gives:
\begin{equation}\label{eq:d_mc}
    d_{mc} = z_{R_{mc}}\frac{1+z_{R_{mc}}/z_{mc}}{1-z_{R_{mc}}/z_{mc}} - z_{mc}
\end{equation}
It also follows that it is only possible to design a mode converter for an input $\tilde{q}_{mc}$ such that $\tan^{-1}(z_{R_{mc}}/z_{mc})<\pi/4$, otherwise it is not possible for the non-focused axis to accumulate $\pi/4$ Gouy phase.

We can now write that the beam spot size at the first and second cylindrical lens, $w_{a}$ and $w_b$ is given by:
\begin{align}
    \frac{\pi w^2_a}{\lambda} &= \mathrm{Re}[ i/\tilde{q}_{mc} ] & \frac{\pi w^2_b}{\lambda} = \mathrm{Re}[ i/(\tilde{q}_{mc}+d_{mc}) ]
\end{align}
To calculate the required focal length in the focusing axis of the two cylindrical lenses, we start by first calculating the required beam parameters after the two cylindrical lenses.
After the first cylindrical lens, the beam has width $w_a$ and propagates a distance $d_{mc}$ and then has width $w_b$.
From the above we know this constraint has two distinct solutions for the beam parameters, one being the non-focused axis, and the other which the beam goes through a waist to give the desired $\pi/2$ difference in Gouy phase.
One can calculate the beam parameter solution after the first lens for the focused axis to be:
\begin{equation}
    \tilde{q}_{a} = z_a + iz_{R_a}
\end{equation}
\begin{equation}
    z_{R_a} = d^2_{mc}\frac{\pi(w^2_a+w_b^2)/\lambda - 2\sqrt{\pi^2 w_a^2 w_b^2/\lambda^2-d^2_{mc}}}
    {\pi^2(w^2_a-w_b^2)^2/\lambda^2 + 4d_{mc}^4}
\end{equation}
\begin{equation}
    z_a = -\frac{d^2_{mc} + \pi z_{R_a}(w^2_a-w_b^2)/\lambda}{2d_{mc}}
\end{equation}

The required focal length of the first cylindrical lens is then the one which transforms $\tilde{q}_{in}$ into $\tilde{q}_a$:
\begin{equation}\label{eq:f_cl1}
    1/f_{CL_1} = \mathrm{Re}[ 1/\tilde{q}_{mc} - 1/\tilde{q}_{a} ]
\end{equation}
Similarly the required focal length of the second cylindrical lens is the one which transforms $\tilde{q}_a+d_{mc}$ to match the $\tilde{q}$ of the non-focused axis beam at the output of the mode converter, leading to:
\begin{equation}\label{eq:f_cl2}
    1/f_{CL_2} = \mathrm{Re}[ 1/(\tilde{q}_{a}+d_{mc}) - 1/(\tilde{q}_{mc}+d_{mc}) ]
\end{equation}

\subsection{Gouy phase telescope}
Fig.~\ref{fig:gouy_phase_telescope} shows the design of the Gouy phase telescope.
The input beam at the first lens $L_1$ is assumed to be at a waist of size $w_{in}$.
The design parameters to give a spot size of $w_{pd}$ at each of the three detectors, which also have relative accumulated Gouy phase separations of $0,60^\circ$ and $120^\circ$ are given by Eq.~\ref{eq:gouy_telescope_1}-\ref{eq:gouy_telescope_3}.
\begin{align}
w' &= \sqrt{\frac{1}{2w^2_{in}-w^2_{pd}}} & d_2 &= \frac{\pi w^2_{pd}}{2\lambda} + \frac{\pi w_{pd}}{2\lambda w'}\label{eq:gouy_telescope_1}\\
d_3 &= \frac{\pi w^2_{pd}(\sqrt{3}+1)}{2\lambda} & &d_4 = \frac{d_3}{2+\sqrt{3}}\label{eq:gouy_telescope_2} \\
f_{L_1} &= \frac{\pi w^2_{in}w_{pd} w'}{\lambda} & f_{L_2}&={-}f_{L_3} = \frac{\pi w^2_{pd}}{\lambda}\label{eq:gouy_telescope_3}
\end{align}
For the sake of brevity, we do not provide the derivation here however one can arrive at these equations using basic Gaussian beam propagation relations and invoking the aforementioned constraints.
The two cylindrical lenses $CL_1$ and $CL_2$ which form the mode converter can in principle be placed at any point before the first detector so long as the instantaneous Gouy phase at the input is $<3\pi/4$.
This is to allow room for the unfocused axes to accumulate $\pi/4$ Gouy phase.
To minimize the propagation length we place the mode converter within the Gouy phase telescope.
The position of the first cylindrical lens $d_0$, is chosen to be 5 degrees of accumulated Gouy phase from the input lens $L_1$.
This adequately spaces $L_1$ from $CL_1$ and reduces the focal power requirements on the cylindrical lenses.
One can then show that $d_0$ is given by Eq.~\ref{eq:psi}-\ref{eq:d_0}.
\begin{equation}\label{eq:psi}
    \psi' = \cot\left(  \tan^{-1}\left( \frac{\pi w^2_{in}}{\lambda f_{L_1}}\right) + \frac{17\pi}{36}\right)
\end{equation}
\begin{equation}\label{eq:d_0}
    d_0 = \frac{\pi w_{in}^2f_{L_1}\left( \lambda f_{L_1} \psi' + \pi w^2_{in}\right)}{\lambda^2 f_{L_1}^2 + \pi^2 w^4_{in}}
\end{equation}
The beam parameter at the input to the mode converter $\tilde{q}_{mc}$ is then given by Eq.~\ref{eq:q_mc}.
\begin{equation}\label{eq:q_mc}
    \tilde{q}_{mc} = \frac{\pi\lambda w^2_{in}f^2_{L_1}(\psi' +i)}{\lambda^2f^2_{L_1} + \pi^2 w^4_{in}}
\end{equation}
The remaining parameters of the mode converter come from the derivation in the prior section.
Substituting Eq.~\ref{eq:q_mc} into the result in Eq.~\ref{eq:d_mc}, we get:
\begin{equation}\label{eq:d_1}
    d_1 = \frac{\pi\lambda w^2_{in}f^2_{L_1}\left( \tan( \tan^{-1}( \psi' ) + \frac{\pi}{4} ) - \psi' \right)}{\lambda^2f^2_{L_1} + \pi^2 w^4_{in}}
\end{equation}
The two remaining parameters, $f_{CL_1}$ and $f_{CL_2}$ are given by substituting Eq.~\ref{eq:q_mc}-~\ref{eq:d_1} into Eq.~\ref{eq:f_cl1}-~\ref{eq:f_cl2}.
These do not have concise analytical forms so we omit from showing them here.

\subsection{Telescope parameters for retrofitting into aLIGO}
As an example design, if we use the telescope input beam size as the current aLIGO detector alignment sensing telescope of $w_{in}=0.8708$~mm and make the spot size on the detectors be $w_{pd}=0.3$~mm one yields the design parameters shown in Table~\ref{tab:aligo_design}.
The total required propagation distance from the telescope to the last detector is $1.582$~m which is manageable on the  existing optical table.
The required lens focal lengths are easily manufactured.
By tweaking the $w_{pd}$ and $w_{in}$ parameters, the placement of the mode converter and exploring the positive/negative solutions for the mode converter it may be possible to find focal lengths which match off-the-shelf optics.

\begin{table}[ht]
\centering
\begin{tabular}{lcr}
\hline
Parameter & Value & unit \\
\hline
$w_{pd}$ & 0.3000 & mm\\
$w_{in}$ & 0.8708 & mm\\
$f_1$ & 562.4 & mm\\
$d_0$ & 145.3 & mm\\
$f_{CL_1}$ & 620.6 & mm\\
$d_1$ & 319.2 & mm\\
$f_{CL_2}$ & 82.1 & mm\\
$d_2$ & 661.9 & mm\\
$d_3$ & 363.0 & mm\\
$f_2$ & 265.7 & mm\\
$d_4$ & 97.3 & mm\\
$f_3$ & -265.7 & mm\\
$d_{total}$ & 1.582 & m\\
\hline
\end{tabular}
\caption{Example design parameters for retrofit of the sensing technique within the aLIGO alignment sensing telescope.}
\label{tab:aligo_design}
\end{table}

\end{appendices}


\bibliographystyle{unsrt}
\bibliography{references}

@article{bond_interferometer_2017,
	title = {Interferometer techniques for gravitational-wave detection},
	volume = {19},
	issn = {1433-8351},
	url = {https://doi.org/10.1007/s41114-016-0002-8},
	doi = {10.1007/s41114-016-0002-8},
	pages = {3},
	number = {1},
	journaltitle = {Living Reviews in Relativity},
	shortjournal = {Living Rev Relativ},
	author = {Bond, Charlotte and Brown, Daniel and Freise, Andreas and Strain, Kenneth A.},
	urldate = {2020-06-29},
	date = {2017-02-17},
	langid = {english},
}

@book{siegman_lasers_1986,
	location = {Mill Valley, California},
	title = {Lasers},
	isbn = {978-0-935702-11-8 978-0-19-855713-5},
	pagetotal = {1283},
	publisher = {University Science Books},
	author = {Siegman, Anthony E.},
	date = {1986},
	langid = {english},
	note = {{OCLC}: 14525287},
}

@article{agatsuma_high-performance_2019,
	title = {High-performance phase camera as a frequency selective laser wavefront sensor for gravitational wave detectors},
	volume = {27},
	rights = {\&\#169; 2019 Optical Society of America},
	issn = {1094-4087},
	url = {https://www.osapublishing.org/oe/abstract.cfm?uri=oe-27-13-18533},
	doi = {10.1364/OE.27.018533},
	pages = {18533--18548},
	number = {13},
	journaltitle = {Optics Express},
	shortjournal = {Opt. Express, {OE}},
	publisher = {Optical Society of America},
	author = {Agatsuma, Kazuhiro and Schaaf, Laura van der and Beuzekom, Martin van and Rabeling, David and Brand, Jo van den},
	urldate = {2020-07-30},
	date = {2019-06-24},
}

@article{erden_accumulated_1997,
	title = {Accumulated Gouy phase shift in Gaussian beam propagation through first-order optical systems},
	volume = {14},
	rights = {\&\#169; 1997 Optical Society of America},
	issn = {1520-8532},
	url = {https://www.osapublishing.org/josaa/abstract.cfm?uri=josaa-14-9-2190},
	doi = {10.1364/JOSAA.14.002190},
	pages = {2190--2194},
	number = {9},
	journaltitle = {{JOSA} A},
	shortjournal = {J. Opt. Soc. Am. A, {JOSAA}},
	publisher = {Optical Society of America},
	author = {Erden, M. Fatih and Ozaktas, Haldun M.},
	urldate = {2020-08-24},
	date = {1997-09-01},
}

@article{ciobanu_mode_2020,
	title = {Mode matching error signals using radio-frequency beam shape modulation},
	volume = {59},
	rights = {\&\#169; 2020 Optical Society of America},
	issn = {2155-3165},
	url = {https://www.osapublishing.org/ao/abstract.cfm?uri=ao-59-31-9884},
	doi = {10.1364/AO.404646},
	pages = {9884--9895},
	number = {31},
	journaltitle = {Applied Optics},
	shortjournal = {Appl. Opt., {AO}},
	publisher = {Optical Society of America},
	author = {Ciobanu, A. A. and Ciobanu, A. A. and Brown, D. D. and Brown, D. D. and Veitch, P. J. and Veitch, P. J. and Ottaway, D. J. and Ottaway, D. J.},
	urldate = {2021-01-14},
	date = {2020-11-01},
}

@article{brooks_overview_2016,
	title = {Overview of Advanced {LIGO} adaptive optics},
	volume = {55},
	rights = {\&\#169; 2016 Optical Society of America},
	issn = {2155-3165},
	url = {https://www.osapublishing.org/ao/abstract.cfm?uri=ao-55-29-8256},
	doi = {10.1364/AO.55.008256},
	pages = {8256--8265},
	number = {29},
	journaltitle = {Applied Optics},
	shortjournal = {Appl. Opt., {AO}},
	publisher = {Optical Society of America},
	author = {Brooks, Aidan F. and Abbott, Benjamin and Arain, Muzammil A. and Ciani, Giacomo and Cole, Ayodele and Grabeel, Greg and Gustafson, Eric and Guido, Chris and Heintze, Matthew and Heptonstall, Alastair and Jacobson, Mindy and Kim, Won and King, Eleanor and Lynch, Alexander and O’Connor, Stephen and Ottaway, David and Mailand, Ken and Mueller, Guido and Munch, Jesper and Sannibale, Virginio and Shao, Zhenhua and Smith, Michael and Veitch, Peter and Vo, Thomas and Vorvick, Cheryl and Willems, Phil},
	urldate = {2021-02-05},
	date = {2016-10-10},
}

@article{muniz_high_2021,
	title = {High frame-rate phase camera for high-resolution wavefront sensing in gravitational-wave detectors},
	volume = {104},
	url = {https://link.aps.org/doi/10.1103/PhysRevD.104.042002},
	doi = {10.1103/PhysRevD.104.042002},
	pages = {042002},
	number = {4},
	journaltitle = {Physical Review D},
	shortjournal = {Phys. Rev. D},
	publisher = {American Physical Society},
	author = {Muñiz, Erik and Srivastava, Varun and Vidyant, Subham and Ballmer, Stefan W.},
	urldate = {2021-09-13},
	date = {2021-08-13},
}

@article{bayer-helms_coupling_1984,
	title = {Coupling coefficients of an incident wave and the modes of a spherical optical resonator in the case of mismatching and misalignment},
	volume = {23},
	rights = {\&\#169; 1984 Optical Society of America},
	issn = {2155-3165},
	url = {https://opg.optica.org/ao/abstract.cfm?uri=ao-23-9-1369},
	doi = {10.1364/AO.23.001369},
	pages = {1369--1380},
	number = {9},
	journaltitle = {Applied Optics},
	shortjournal = {Appl. Opt., {AO}},
	publisher = {Optical Society of America},
	author = {Bayer-Helms, F.},
	urldate = {2022-02-04},
	date = {1984-05-01},
}

@article{vinet_special_2009,
	title = {On Special Optical Modes and Thermal Issues in Advanced Gravitational Wave Interferometric Detectors},
	volume = {12},
	issn = {1433-8351},
	url = {https://doi.org/10.12942/lrr-2009-5},
	doi = {10.12942/lrr-2009-5},
	pages = {5},
	number = {1},
	journaltitle = {Living Reviews in Relativity},
	shortjournal = {Living Rev. Relativ.},
	author = {Vinet, Jean-Yves},
	urldate = {2022-11-12},
	date = {2009-12-01},
	langid = {english},
}

@thesis{lawrence_active_2003,
	title = {Active wavefront correction in laser interferometric gravitational wave detectors},
	rights = {M.I.T. theses are protected by copyright. They may be viewed from this source for any purpose, but reproduction or distribution in any format is prohibited without written permission. See provided {URL} for inquiries about permission.},
	url = {https://dspace.mit.edu/handle/1721.1/29308},
	institution = {Massachusetts Institute of Technology},
	type = {Thesis},
	author = {Lawrence, Ryan Christopher},
	urldate = {2023-11-06},
	date = {2003},
	note = {Accepted: 2005-10-14T19:46:30Z},
}

@article{hello_analytical_1990,
	title = {Analytical models of thermal aberrations in massive mirrors heated by high power laser beams},
	volume = {51},
	issn = {0302-0738, 2777-3396},
	url = {http://dx.doi.org/10.1051/jphys:0199000510120126700},
	doi = {10.1051/jphys:0199000510120126700},
	pages = {1267--1282},
	number = {12},
	journaltitle = {Journal de Physique},
	shortjournal = {J. Phys. France},
	publisher = {Société Française de Physique},
	author = {Hello, Patrice and Vinet, Jean-Yves},
	urldate = {2023-11-09},
	date = {1990-06-01},
	langid = {english},
}

@article{hello_analytical_1990-1,
	title = {Analytical models of transient thermoelastic deformations of mirrors heated by high power cw laser beams},
	volume = {51},
	issn = {0302-0738, 2777-3396},
	url = {http://dx.doi.org/10.1051/jphys:0199000510200224300},
	doi = {10.1051/jphys:0199000510200224300},
	pages = {2243--2261},
	number = {20},
	journaltitle = {Journal de Physique},
	shortjournal = {J. Phys. France},
	publisher = {Société Française de Physique},
	author = {Hello, Patrice and Vinet, Jean-Yves},
	urldate = {2023-11-09},
	date = {1990-10-01},
	langid = {english},
}

@article{drever_laser_1983,
	title = {Laser phase and frequency stabilization using an optical resonator},
	volume = {31},
	issn = {1432-0649},
	url = {https://doi.org/10.1007/BF00702605},
	doi = {10.1007/BF00702605},
	pages = {97--105},
	number = {2},
	journaltitle = {Applied Physics B},
	shortjournal = {Appl. Phys. B},
	author = {Drever, R. W. P. and Hall, J. L. and Kowalski, F. V. and Hough, J. and Ford, G. M. and Munley, A. J. and Ward, H.},
	urldate = {2023-12-14},
	date = {1983-06-01},
	langid = {english},
}

@article{mueller_determination_2000,
	title = {Determination and optimization of mode matching into optical cavities by heterodyne detection},
	volume = {25},
	rights = {© 2000 Optical Society of America},
	issn = {1539-4794},
	url = {https://opg.optica.org/ol/abstract.cfm?uri=ol-25-4-266},
	doi = {10.1364/OL.25.000266},
	pages = {266--268},
	number = {4},
	journaltitle = {Optics Letters},
	shortjournal = {Opt. Lett., {OL}},
	publisher = {Optica Publishing Group},
	author = {Mueller, Guido and Shu, Qi-ze and Adhikari, Rana and Tanner, D. B. and Reitze, David and Sigg, Daniel and Mavalvala, Nergis and Camp, Jordan},
	urldate = {2023-12-19},
	date = {2000-02-15},
}

@article{morrison_automatic_1994,
	title = {Automatic alignment of optical interferometers},
	volume = {33},
	rights = {© 1994 Optical Society of America},
	issn = {2155-3165},
	url = {https://opg.optica.org/ao/abstract.cfm?uri=ao-33-22-5041},
	doi = {10.1364/AO.33.005041},
	pages = {5041--5049},
	number = {22},
	journaltitle = {Applied Optics},
	shortjournal = {Appl. Opt., {AO}},
	publisher = {Optica Publishing Group},
	author = {Morrison, Euan and Meers, Brian J. and Robertson, David I. and Ward, Henry},
	urldate = {2024-02-06},
	date = {1994-08-01},
}

@article{black_introduction_2001,
	title = {An introduction to Pound–Drever–Hall laser frequency stabilization},
	volume = {69},
	issn = {0002-9505},
	url = {https://doi.org/10.1119/1.1286663},
	doi = {10.1119/1.1286663},
	pages = {79--87},
	number = {1},
	journaltitle = {American Journal of Physics},
	shortjournal = {American Journal of Physics},
	author = {Black, Eric D.},
	urldate = {2024-02-08},
	date = {2001-01-01},
}

@article{goodwin-jones_transverse_2024,
	title = {Transverse mode control in quantum enhanced interferometers: a review and recommendations for a new generation},
	volume = {11},
	rights = {\&\#169; 2024 Optica Publishing Group},
	issn = {2334-2536},
	url = {https://opg.optica.org/optica/abstract.cfm?uri=optica-11-2-273},
	doi = {10.1364/OPTICA.511924},
	shorttitle = {Transverse mode control in quantum enhanced interferometers},
	pages = {273--290},
	number = {2},
	journaltitle = {Optica},
	shortjournal = {Optica, {OPTICA}},
	publisher = {Optica Publishing Group},
	author = {Goodwin-Jones, Aaron W. and Cabrita, Ricardo and Korobko, Mikhail and Beuzekom, Martin Van and Brown, Daniel D. and Fafone, Viviana and Heijningen, Joris Van and Rocchi, Alessio and Schiworski, Mitchell G. and Tacca, Matteo},
	urldate = {2024-02-26},
	date = {2024-02-20},
}

@misc{cao_optical_2019,
	title = {An optical lock-in camera for advanced gravitational wave interferometers},
	url = {http://arxiv.org/abs/1907.05224},
	number = {{arXiv}:1907.05224},
	publisher = {{arXiv}},
	author = {Cao, Huy Tuong and Brown, Daniel D. and Veitch, Peter and Ottaway, David J.},
	urldate = {2024-03-13},
	date = {2019-07-10},
	langid = {english},
	eprinttype = {arxiv},
	eprint = {1907.05224 [astro-ph, physics:physics]},
}

@article{simonin_negative_2016,
	title = {Negative ion source development for a photoneutralization based neutral beam system for future fusion reactors},
	volume = {18},
	issn = {1367-2630},
	url = {https://dx.doi.org/10.1088/1367-2630/18/12/125005},
	doi = {10.1088/1367-2630/18/12/125005},
	pages = {125005},
	number = {12},
	journaltitle = {New Journal of Physics},
	shortjournal = {New J. Phys.},
	publisher = {{IOP} Publishing},
	author = {Simonin, A. and Agnello, R. and Bechu, S. and Bernard, J. M. and Blondel, C. and Boeuf, J. P. and Bresteau, D. and Cartry, G. and Chaibi, W. and Drag, C. and Duval, B. P. and Esch, H. P. L. de and Fubiani, G. and Furno, I. and Grand, C. and Guittienne, Ph and Howling, A. and Jacquier, R. and Marini, C. and Morgal, I.},
	urldate = {2024-11-25},
	date = {2016-12},
	langid = {english},
}

@article{magana-sandoval_sensing_2019,
	title = {Sensing optical cavity mismatch with a mode-converter and quadrant photodiode},
	volume = {100},
	url = {https://link.aps.org/doi/10.1103/PhysRevD.100.102001},
	doi = {10.1103/PhysRevD.100.102001},
	pages = {102001},
	number = {10},
	journaltitle = {Physical Review D},
	shortjournal = {Phys. Rev. D},
	publisher = {American Physical Society},
	author = {Magaña-Sandoval, Fabian and Vo, Thomas and Vander-Hyde, Daniel and Sanders, J. R. and Ballmer, Stefan W.},
	urldate = {2025-01-10},
	date = {2019-11-18},
}

@article{oneil_mode_2000,
	title = {Mode transformations in terms of the constituent Hermite–Gaussian or Laguerre–Gaussian modes and the variable-phase mode converter},
	volume = {181},
	issn = {0030-4018},
	url = {https://www.sciencedirect.com/science/article/pii/S0030401800007367},
	doi = {10.1016/S0030-4018(00)00736-7},
	pages = {35--45},
	number = {1},
	journaltitle = {Optics Communications},
	shortjournal = {Optics Communications},
	author = {O'Neil, Anna T. and Courtial, Johannes},
	urldate = {2025-01-21},
	date = {2000-07-01},
}

@article{anderson_alignment_1984,
	title = {Alignment of resonant optical cavities},
	volume = {23},
	rights = {© 1984 Optical Society of America},
	issn = {2155-3165},
	url = {https://opg.optica.org/ao/abstract.cfm?uri=ao-23-17-2944},
	doi = {10.1364/AO.23.002944},
	pages = {2944--2949},
	number = {17},
	journaltitle = {Applied Optics},
	shortjournal = {Appl. Opt., {AO}},
	publisher = {Optica Publishing Group},
	author = {Anderson, Dana Z.},
	urldate = {2025-02-05},
	date = {1984-09-01},
}

@misc{bond_analytical_2016,
	title = {Analytical calculation of Hermite-Gauss and Laguerre-Gauss modes on a bullseye photodiode},
	url = {http://arxiv.org/abs/1606.01057},
	doi = {10.48550/arXiv.1606.01057},
	number = {{arXiv}:1606.01057},
	publisher = {{arXiv}},
	author = {Bond, Charlotte and Fulda, Paul and Freise, Andreas},
	urldate = {2025-10-21},
	date = {2016-06-03},
	eprinttype = {arxiv},
	eprint = {1606.01057 [physics]},
}

@article{beijersbergen_astigmatic_nodate,
	title = {Astigmatic laser mode converters and transfer of orbital angular momentum},
	author = {Beijersbergen, M W and Allen, L},
	langid = {english},
}

@thesis{schiworski_development_2024,
	title = {Development and application of phase cameras for advanced gravitational wave detectors},
	url = {https://hdl.handle.net/2440/145015},
	institution = {The University of Adelaide},
	type = {phdthesis},
	author = {Schiworski, Mitchell G.},
	date = {2024},
}

@article{martens_design_2022,
	title = {Design of the optical system for the gamma factory proof of principle experiment at the {CERN} Super Proton Synchrotron},
	volume = {25},
	doi = {10.1103/PhysRevAccelBeams.25.101601},
	number = {10},
	journaltitle = {Physical Review Accelerators and Beams},
	shortjournal = {Phys. Rev. Accel. Beams},
	author = {Martens, Aurélien},
	date = {2022},
}

@misc{kuns_squeezed_2026,
	title = {Squeezed state degradations due to mode mismatch and thermal aberrations in gravitational wave detectors},
	url = {http://arxiv.org/abs/2604.23835},
	doi = {10.48550/arXiv.2604.23835},
	number = {{arXiv}:2604.23835},
	publisher = {{arXiv}},
	author = {Kuns, Kevin and Brown, Daniel},
	urldate = {2026-04-28},
	date = {2026-04-26},
	eprinttype = {arxiv},
	eprint = {2604.23835 [physics]},
}

@article{mcculler_ligos_2021,
	title = {{LIGO}'s quantum response to squeezed states},
	volume = {104},
	url = {https://link.aps.org/doi/10.1103/PhysRevD.104.062006},
	doi = {10.1103/PhysRevD.104.062006},
	pages = {062006},
	number = {6},
	journaltitle = {Physical Review D},
	shortjournal = {Phys. Rev. D},
	publisher = {American Physical Society},
	author = {{McCuller}, L. and Dwyer, S. E. and Green, A. C. and Yu, Haocun and Kuns, K. and Barsotti, L. and Blair, C. D. and Brown, D. D. and Effler, A. and Evans, M. and Fernandez-Galiana, A. and Fritschel, P. and Frolov, V. V. and Kijbunchoo, N. and Mansell, G. L. and Matichard, F. and Mavalvala, N. and {McClelland}, D. E. and {McRae}, T. and Mullavey, A. and Sigg, D. and Slagmolen, B. J. J. and Tse, M. and Vo, T. and Ward, R. L. and Whittle, C. and Abbott, R. and Adams, C. and Adhikari, R. X. and Ananyeva, A. and Appert, S. and Arai, K. and Areeda, J. S. and Asali, Y. and Aston, S. M. and Austin, C. and Baer, A. M. and Ball, M. and Ballmer, S. W. and Banagiri, S. and Barker, D. and Bartlett, J. and Berger, B. K. and Betzwieser, J. and Bhattacharjee, D. and Billingsley, G. and Biscans, S. and Blair, R. M. and Bode, N. and Booker, P. and Bork, R. and Bramley, A. and Brooks, A. F. and Buikema, A. and Cahillane, C. and Cannon, K. C. and Chen, X. and Ciobanu, A. A. and Clara, F. and Compton, C. M. and Cooper, S. J. and Corley, K. R. and Countryman, S. T. and Covas, P. B. and Coyne, D. C. and Datrier, L. E. H. and Davis, D. and Di Fronzo, C. and Dooley, K. L. and Driggers, J. C. and Etzel, T. and Evans, T. M. and Feicht, J. and Fulda, P. and Fyffe, M. and Giaime, J. A. and Giardina, K. D. and Godwin, P. and Goetz, E. and Gras, S. and Gray, C. and Gray, R. and Gustafson, E. K. and Gustafson, R. and Hanks, J. and Hanson, J. and Hardwick, T. and Hasskew, R. K. and Heintze, M. C. and Helmling-Cornell, A. F. and Holland, N. A. and Jones, J. D. and Kandhasamy, S. and Karki, S. and Kasprzack, M. and Kawabe, K. and King, P. J. and Kissel, J. S. and Kumar, Rahul and Landry, M. and Lane, B. B. and Lantz, B. and Laxen, M. and Lecoeuche, Y. K. and Leviton, J. and Liu, J. and Lormand, M. and Lundgren, A. P. and Macas, R. and {MacInnis}, M. and Macleod, D. M. and Márka, S. and Márka, Z. and Martynov, D. V. and Mason, K. and Massinger, T. J. and {McCarthy}, R. and {McCormick}, S. and {McIver}, J. and Mendell, G. and Merfeld, K. and Merilh, E. L. and Meylahn, F. and Mistry, T. and Mittleman, R. and Moreno, G. and Mow-Lowry, C. M. and Mozzon, S. and Nelson, T. J. N. and Nguyen, P. and Nuttall, L. K. and Oberling, J. and Oram, Richard J. and Osthelder, C. and Ottaway, D. J. and Overmier, H. and Palamos, J. R. and Parker, W. and Payne, E. and Pele, A. and Penhorwood, R. and Perez, C. J. and Pirello, M. and Radkins, H. and Ramirez, K. E. and Richardson, J. W. and Riles, K. and Robertson, N. A. and Rollins, J. G. and Romel, C. L. and Romie, J. H. and Ross, M. P. and Ryan, K. and Sadecki, T. and Sanchez, E. J. and Sanchez, L. E. and Saravanan, T. R. and Savage, R. L. and Schaetzl, D. and Schnabel, R. and Schofield, R. M. S. and Schwartz, E. and Sellers, D. and Shaffer, T. and Smith, J. R. and Soni, S. and Sorazu, B. and Spencer, A. P. and Strain, K. A. and Sun, L. and Szczepańczyk, M. J. and Thomas, M. and Thomas, P. and Thorne, K. A. and Toland, K. and Torrie, C. I. and Traylor, G. and Urban, A. L. and Vajente, G. and Valdes, G. and Vander-Hyde, D. C. and Veitch, P. J. and Venkateswara, K. and Venugopalan, G. and Viets, A. D. and Vorvick, C. and Wade, M. and Warner, J. and Weaver, B. and Weiss, R. and Willke, B. and Wipf, C. C. and Xiao, L. and Yamamoto, H. and Yu, Hang and Zhang, L. and Zucker, M. E. and Zweizig, J.},
	urldate = {2026-04-28},
	date = {2021-09-13},
}

@article{pupeza_compact_2013,
	title = {Compact high-repetition-rate source of coherent 100 {eV} radiation},
	volume = {7},
	rights = {2013 Springer Nature Limited},
	issn = {1749-4893},
	url = {https://www.nature.com/articles/nphoton.2013.156},
	doi = {10.1038/nphoton.2013.156},
	pages = {608--612},
	number = {8},
	journaltitle = {Nature Photonics},
	shortjournal = {Nature Photon},
	publisher = {Nature Publishing Group},
	author = {Pupeza, I. and Holzberger, S. and Eidam, T. and Carstens, H. and Esser, D. and Weitenberg, J. and Rußbüldt, P. and Rauschenberger, J. and Limpert, J. and Udem, Th and Tünnermann, A. and Hänsch, T. W. and Apolonski, A. and Krausz, F. and Fill, E.},
	urldate = {2026-04-28},
	date = {2013-08},
	langid = {english},
}

@article{jacquet_first_2024,
	title = {First production of X-rays at the {ThomX} high-intensity Compton source},
	volume = {139},
	issn = {2190-5444},
	url = {https://doi.org/10.1140/epjp/s13360-024-05186-z},
	doi = {10.1140/epjp/s13360-024-05186-z},
	pages = {459},
	number = {5},
	journaltitle = {The European Physical Journal Plus},
	shortjournal = {Eur. Phys. J.  Plus},
	author = {Jacquet, Marie and Alexandre, Patrick and Alkadi, Muath and Alves, Manuel and Amer, Manar and Amoudry, Loic and Auguste, Didier and Babigeon, Jean-Luc and Balcou, Philippe and Baltazar, Michel and Benabderrahmane, Chamseddine and el Fekih, Rachid Ben and Benoit, Alain and Berteaud, Philippe and Biagini, Marica and Blin, Alexandre and Bobault, Sébastien and Bonanzingamarco, Marco and Bonenfant, Jean and Bonis, Julien and Bouanani, Yazid and Bouaziz, Said and Bouvet, François and Bravin, Alberto and Bruni, Christelle and Bruyere, Cyril and Bzyl, Harold and Cassinari, Lodovico and Cassou, Kevin and Cayla, Jean-Noël and Chabaud, Thomas and Chaikovska, Iryna and Chance, Sophie and Chapelle, Christophe and Chaumat, Vincent and Chiche, Ronic and Cobessi, Alain and Cormier, Eric and Cornebise, Patrick and Couprie, Marie-Emmanuelle and Cuoq, Renaud and Dalifard, Olivier and Degallaix, Jérôme and Delerue, Nicolas and Del Net, William and Diaz, Antonio and Dietrich, Yannick and Diop, Massamba and Dorkel, Remy and Douillet, Denis and Drebot, Illya and Dugal, Jean-Phillipe and Dupraz, Kevin and Dupuy, Eric and El Ajjouri, Moussa and El Kamchi, Noureddine and El Khaldi, Mohamed and Elleaume, Hélène and Ergenlik, Ezgi and Estève, François and Favier, Pierre and Fernandez, Marco and Gamelin, Alexis and Garaut, Jean-François and Garolfi, Luca and Gauron, Philippe and Gauthier, Frédéric and Girault, Pascal and Gonnin, Alexandre and Grasset, Denis and Guerard, Eric and Guler, Hayg and Haissinski, Jacques and Hazemann, Jean-Louis and Helder, Dias and Herbeaux, Christian and Herry, Emmanuel and Hodeau, Jean-Louis and Horodynski, Jean-Michel and Hubert, Nicolas and Iaquaniello, Gregory and Jacquet, Philippe and Jeantet, Philippe and Jehanno, Didier and Jules, Eric and Kapoujyan, Grigor and Kubytskyi, Viacheslav and Labat, Marie and Labaye, François and Lacipière, Jerome and Lacroix, Mickaël and Lahéra, Eric and Langlet, Marc and Lebarillec, Titouan and Ledu, Jean-François and Le Duc, Géraldine and Le Guidec, Damien and Leluan, Bruno and Lepercq, Pierre and Lestrade, Alain and Letellier-cohen, Frédéric and Letrésor, Antoine and Lhermite, Jérôme and Liu, Xing and Lopes, Robert and Loulergue, Alexandre and Louvet, Marc and Mageur, Christophe and Marchand, Patrick and Marie, Rodolphe and Marrucho, Jean-Claude and Marteau, Fabrice and Martens, Aurélien and Mercadier, Gabriel and Mercier, Bruno and Michel, Christophe and Mistretta, Eric and Monard, Hugues and Moutardier, Alexandre and Muller, Didier and Nadji, Amor and Nadolski, Laurent and Nagaoka, Ryutaro and Neveu, Olivier and Nutarelli, Daniele and Omeich, Maher and Pedeau, Dominique and Peinaud, Yann and Perroux, Gilles and Pérus, Antoine and Petit, Sylvain and Petrilli, Yannick and Pichet, Marc and Pieyre, Bernard and Pinard, Laurent and Plaige, Eric and Pollina, Jean Pierre and Prévost, Christophe and Proux, Olivier and Ribeiro, Fernand and Robert, Pierre and Ros, Manuel and Roulet, Thomas and Roux, Raphael and Roy, Emmanuel and Rudnicky, Philippe and Salvia, Julien and Sebdaoui, Mourad and Sierra, Serge and Soskov, Viktor and Sreedharan, Rajesh and Susini, Jean and Taurigna-Quéré, Monique and Trochet, Stéphane and Vallerand, Cynthia and Variola, Alessandro and Veteran, José and Vitez, Olivier and Walter, Philippe and Wicek, François and Wurth, Sébastien and Zomer, Fabian},
	urldate = {2026-04-28},
	date = {2024-05-30},
	langid = {english},
}

@article{brown_differential_2021,
	title = {Differential wavefront sensing and control using radio-frequency optical demodulation},
	volume = {29},
	rights = {© 2021 Optical Society of America},
	issn = {1094-4087},
	url = {https://opg.optica.org/oe/abstract.cfm?uri=oe-29-11-15995},
	doi = {10.1364/OE.425590},
	pages = {15995--16006},
	number = {11},
	journaltitle = {Optics Express},
	shortjournal = {Opt. Express, {OE}},
	publisher = {Optica Publishing Group},
	author = {Brown, Daniel and Cao, Huy Tuong and Ciobanu, Alexei and Veitch, Peter and Ottaway, David},
	urldate = {2026-06-02},
	date = {2021-05-24},
}

@article{goodwin-jones_single_2023,
	title = {Single and coupled cavity mode sensing schemes using a diagnostic field},
	volume = {31},
	rights = {© 2023 Optica Publishing Group},
	issn = {1094-4087},
	url = {https://opg.optica.org/oe/abstract.cfm?uri=oe-31-21-35068},
	doi = {10.1364/OE.502911},
	pages = {35068--35085},
	number = {21},
	journaltitle = {Optics Express},
	shortjournal = {Opt. Express, {OE}},
	publisher = {Optica Publishing Group},
	author = {Goodwin-Jones, Aaron W. and Zhu, Haochen and Blair, Carl and Brown, Daniel D. and Heijningen, Joris van and Ju, Li and Zhao, Chunnong},
	urldate = {2026-06-22},
	date = {2023-10-09},
}

@article{rocchi_thermal_2012,
	title = {Thermal effects and their compensation in Advanced Virgo},
	volume = {363},
	issn = {1742-6596},
	url = {https://doi.org/10.1088/1742-6596/363/1/012016},
	doi = {10.1088/1742-6596/363/1/012016},
	pages = {012016},
	number = {1},
	journaltitle = {Journal of Physics: Conference Series},
	shortjournal = {J. Phys.: Conf. Ser.},
	author = {Rocchi, A and Coccia, E and Fafone, V and Malvezzi, V and Minenkov, Y and Sperandio, L},
	urldate = {2026-07-16},
	date = {2012-06},
	langid = {english},
}

@report{kawabe_orientation_2006,
	title = {Orientation of Quadrant Diode for Wave Front Sensing},
	url = {https://dcc.ligo.org/LIGO-T060035},
	number = {T060035},
	institution = {{LIGO} {DCC}},
	author = {Kawabe, Keita},
	date = {2006},
}

@software{brown_2025_12662017,
  author       = {Brown, Daniel David and
                  Freise, Andreas and
                  Cao, Huy Tuong and
                  Ciobanu, Alexei and
                  Gobeil, Jeremie and
                  Green, Anna and
                  Hapke, Paul and
                  Jones, Philip and
                  van der Kolk, Miron and
                  Kuns, Kevin and
                  Leavey, Sean and
                  Perry, Jonathan Warren and
                  Rowlinson, Samuel and
                  Sallé, Mischa},
  title        = {FINESSE},
  month        = mar,
  year         = 2025,
  publisher    = {Gitlab},
  version      = {3.0a32},
  doi          = {10.5281/zenodo.12662017},
  url          = {https://doi.org/10.5281/zenodo.12662017},
}

\end{document}